\documentclass[sn-mathphys-num]{sn-jnl}
\usepackage{graphicx}%
\usepackage{multirow}%
\usepackage{amsmath,amssymb,amsfonts}%
\usepackage{amsthm}%
\usepackage{mathrsfs}%
\usepackage[title]{appendix}%
\usepackage{xcolor}%
\usepackage{textcomp}%
\usepackage{manyfoot}%
\usepackage{booktabs}%
\usepackage{algorithm}%
\usepackage{algorithmicx}%
\usepackage{algpseudocode}%
\usepackage{listings}%
\usepackage{svg}%
\usepackage{subcaption}%
\theoremstyle{thmstyleone}%
\usepackage{subcaption}
\usepackage{lineno}
\theoremstyle{thmstyletwo}%

\theoremstyle{thmstylethree}%

\begin{document}

\title[Machine learning for radiometer calibration]{Radiometer Calibration using Machine Learning}

\author*[1,2]{\fnm{S. A. K.} \sur{Leeney}}\email{sakl2@cam.ac.uk}
\author[1,2]{\fnm{H. T. J.} \sur{Bevins}} 
\author[1,2]{\fnm{E.} \sur{de Lera Acedo}}
\author[2,3]{\fnm{W. J.} \sur{Handley}}
\author[1,2]{\fnm{C.} \sur{Kirkham}}
\author[1,2]{\fnm{R. S.} \sur{Patel}}
\author[1,4, 19]{\fnm{J.} \sur{Zhu}}
\author[1,2]{\fnm{D.} \sur{Molnar}}
\author[1,2]{\fnm{J.} \sur{Cumner}}
\author[1,2]{\fnm{D.} \sur{Anstey}}
\author[1,2]{\fnm{K.} \sur{Artuc}}
\author[5]{\fnm{G.} \sur{Bernardi}}
\author[6,7]{\fnm{M.} \sur{Bucher}}
\author[1]{\fnm{S.} \sur{Carey}}
\author[8]{\fnm{J.} \sur{Cavillot}}
\author[9]{\fnm{R.} \sur{Chiello}}
\author[7]{\fnm{W.} \sur{Croukamp}}
\author[7]{\fnm{D. I. L.} \sur{de Villiers}}
\author[1]{\fnm{J. A.} \sur{Ely}}
\author[2,3]{\fnm{A.} \sur{Fialkov}}
\author[1,2]{\fnm{T.} \sur{Gessey-Jones}}
\author[10]{\fnm{G.} \sur{Kulkarni}}
\author[11]{\fnm{A.} \sur{Magro}}
\author[12]{\fnm{P. D.} \sur{Meerburg}}
\author[1,2]{\fnm{S.} \sur{Mittal}}
\author[1,2]{\fnm{J. H. N.} \sur{Pattison}}
\author[7,13]{\fnm{S.} \sur{Pegwal}}
\author[7]{\fnm{C. M.} \sur{Pieterse}}
\author[14]{\fnm{J. R.} \sur{Pritchard}}
\author[15]{\fnm{E.} \sur{Puchwein}}
\author[1]{\fnm{N.} \sur{Razavi-Ghods}}
\author[1]{\fnm{I. L. V.} \sur{Roque}}
\author[12]{\fnm{A.} \sur{Saxena}}
\author[1,2]{\fnm{K. H.} \sur{Scheutwinkel}}
\author[1]{\fnm{P.} \sur{Scott}}
\author[1,2]{\fnm{E.} \sur{Shen}}
\author[16]{\fnm{P. H.} \sur{Sims}}
\author[17,18]{\fnm{M.} \sur{Spinelli}}

\affil[1]{\orgdiv{Astrophysics Group, Cavendish Laboratory}, \orgname{University of Cambridge}, \orgaddress{\street{J. J. Thomson Avenue}, \city{Cambridge}, \postcode{CB3 0HE}, \country{UK}}}

\affil[2]{\orgdiv{Kavli Institute for Cosmology in Cambridge}, \orgname{University of Cambridge}, \orgaddress{\street{Madingley Road}, \city{Cambridge}, \postcode{CB3 0HA}, \country{UK}}}

\affil[3]{\orgdiv{Institute of Astronomy}, \orgname{University of Cambridge}, \orgaddress{\street{Madingley Road}, \city{Cambridge}, \postcode{CB3 0HA}, \country{UK}}}

\affil[4]{\orgdiv{National Astronomical Observatory}, \orgname{Chinese Academy of Science}, \orgaddress{Beijing, 100101, \country{China}}}

\affil[5]{\orgdiv{INAF-Istituto di Radio Astronomia}, \orgaddress{\street{Via Gobetti 101}, \city{Bologna}, \postcode{40129}, \country{Italy}}}

\affil[6]{\orgdiv{Laboratoire AstroParticule et Cosmologie}, \orgname{Université Paris-Cité}, \orgaddress{\street{10 Rue Alice Domon et Léonie Duquet}, \city{Paris}, \postcode{75013}, \country{France}}}

\affil[7]{\orgdiv{Department of Electrical and Electronic Engineering}, \orgname{Stellenbosch University}, \orgaddress{\city{Stellenbosch}, \postcode{7602}, \country{South Africa}}}

\affil[8]{\orgdiv{Antenna Group}, \orgname{Université catholique de Louvain}, \orgaddress{\city{Louvain-la-Neuve}, \postcode{1348}, \country{Belgium}}}

\affil[9]{\orgdiv{Physics Department}, \orgname{University of Oxford}, \orgaddress{\street{Parks Road}, \city{Oxford}, \postcode{OX1 3PU}, \country{UK}}}

\affil[10]{\orgdiv{Department of Theoretical Physics}, \orgname{Tata Institute of Fundamental Research}, \orgaddress{\street{Homi Bhabha Road}, \city{Mumbai}, \postcode{400005}, \country{India}}}

\affil[11]{\orgdiv{Institute of Space Sciences and Astronomy}, \orgname{University of Malta}, \orgaddress{\street{Msida}, \city{Malta}, \postcode{MSD 2080}, \country{Malta}}}

\affil[12]{\orgdiv{Faculty of Science and Engineering}, \orgname{University of Groningen}, \orgaddress{\street{Nijenborgh 4}, \city{Groningen}, \postcode{9747 AG}, \country{Netherlands}}}

\affil[13]{\orgdiv{South African Radio Astronomy Observatory}, \orgaddress{\street{Black River Park, 2 Fir Street, Observatory}, \city{Cape Town}, \postcode{7925}, \country{South Africa}}}

\affil[14]{\orgdiv{Department of Physics}, \orgname{Imperial College London}, \orgaddress{\street{South Kensington Campus}, \city{London}, \postcode{SW7 2AZ}, \country{UK}}}

\affil[15]{\orgdiv{Leibniz Institute for Astrophysics}, \orgaddress{\street{An der Sternwarte 16}, \city{Potsdam}, \postcode{14482}, \country{Germany}}}

\affil[16]{\orgdiv{Department of Physics}, \orgname{Arizona State University}, \orgaddress{\street{781 South Terrace Rd}, \city{Tempe}, \postcode{6004}, \country{US}}}

\affil[17]{\orgdiv{Observatoire de la Côte d’Azur}, \city{Nice}, \country{France}}

\affil[18]{\orgdiv{Department of Physics and Astronomy}, \orgname{University of the Western Cape}, \orgaddress{\street{Robert Sobukhwe Road}, \city{Bellville}, \postcode{7535}, \country{South Africa}}}
\affil[19]{\orgdiv{University of Chinese Academy of Sciences}, \orgaddress{Beijing, 100049, \country{China}}}

\abstract{Radiometers are crucial instruments in radio astronomy, forming the primary component of nearly all radio telescopes. They measure the intensity of electromagnetic radiation, converting this radiation into electrical signals. A radiometer's primary components are an antenna and a Low Noise Amplifier (LNA), which is the core of the ``receiver'' chain. Instrumental effects introduced by the receiver are typically corrected or removed during calibration. However, impedance mismatches between the antenna and receiver can introduce unwanted signal reflections and distortions. Traditional calibration methods, such as Dicke switching, alternate the receiver input between the antenna and a well-characterised reference source to mitigate errors by comparison. Recent advances in Machine Learning (ML) offer promising alternatives. Neural networks, which are trained using known signal sources, provide a powerful means to model and calibrate complex systems where traditional analytical approaches struggle. These methods are especially relevant for detecting the faint sky-averaged 21-cm signal from atomic hydrogen at high redshifts. This is one of the main challenges in observational Cosmology today. Here, for the first time, we introduce and test a machine learning-based calibration framework capable of achieving the precision required for radiometric experiments aiming to detect the 21-cm line.}

\maketitle

\section{Introduction}\label{sec:intro}
Radiometers have been integral to the field of radio astronomy since its inception. They measure the intensity of incoming electromagnetic radiation within a specific frequency band, producing a proportional electrical signal at their output. Typically, a radiometer consists of a radio antenna that captures the radiant energy and an amplifying device such as a Low Noise Amplifier (LNA) which, in combination with several other components described in Section~\ref{sec:radiometrcalibration}, is commonly referred to as the ``receiver''. The incident electromagnetic radiation induces a voltage across the antenna, which can be interpreted as Johnson-Nyquist noise representative of the sky temperature $( T_{\text{sky}})$~\cite{nyquist1928thermal}. For cosmological applications, particularly those aiming to detect faint signals like the sky-averaged 21-cm line of neutral hydrogen, this initial voltage signal is too weak to be directly interpreted. The signal must therefore be amplified by the receiver before it can be measured. This amplification modifies the signal non-linearly across frequencies, introducing a multiplicative gain factor, $g$, and an additive receiver noise term, $T_{\text{rec}}$.

It is typically very difficult to know these two values a priori, which motivates the need for external references or calibration sources to constrain them. The relationship between the measured output power and the input temperature from a source can be modelled as

\begin{equation}\label{eq:base_t} \mathcal{P}_{\text{src}}^{\text{out}} = g M(T^{\text{in}}_{\text{sky}} + T_{\text{rec}}), \end{equation}

where M (defined in Section~\ref{sec:methods}) is the mismatch factor. This equation highlights how the measured power depends on both the intrinsic properties of the source and the receiver’s internal noise and impedance mismatch.

\begin{figure}[h]
	\centering
	\includegraphics[width=0.8\textwidth]{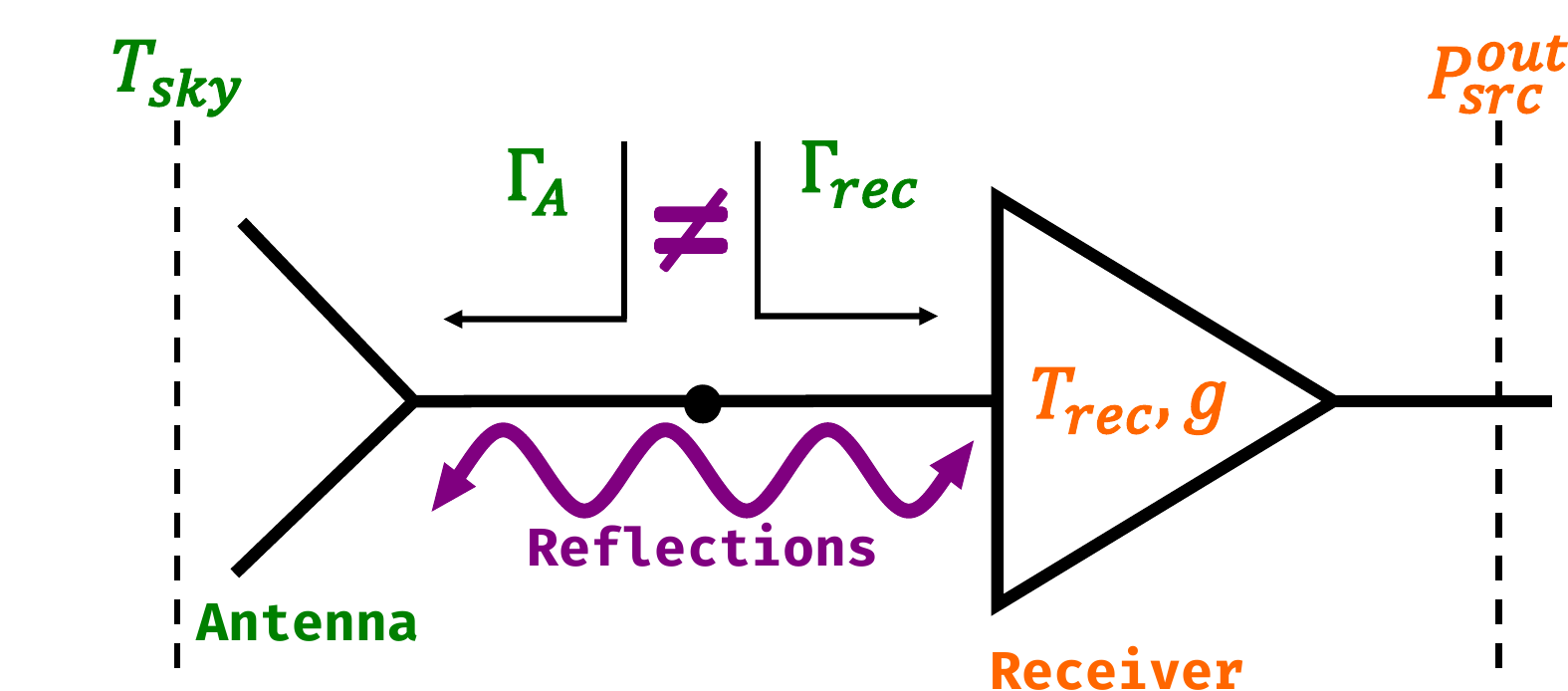}
	\caption{A schematic overview explaining the core challenge in radiometer calibration. An impedance mismatch between the receiver and the source (usually a radio antenna) causes standing waves between the receiver and source. This adds power to the output signal and must be corrected for via calibration. $\Gamma_{A}$ and $\Gamma_{\text{rec}}$ represent the reflection coefficient of the antenna and the receiver and $P_{\text{src}}^{\text{out}}$  is the measured power. Throughout the paper, green and orange represent ``inputs'' and ``outputs'', which will be described in detail in Section~\ref{sec:radiometrcalibration}.}
	\label{fig:receiver_intro}
\end{figure}

Mismatches between the impedances of the antenna and receiver result in partial reflections of both the power incoming from the sky and the noise power generated by the amplifier, shown in Figure~\ref{fig:receiver_intro}. These reflections can lead to standing waves and other spectral distortions in the measured output power that complicate the interpretation of the signal. These are typically referred to as ``systematics''.

Temporal instabilities in the receiver electronics amplify these challenges, introducing perturbations in the signal that are hard to calibrate. Although integrating time and frequency data can help mitigate random or radiometric noise, eliminating the systematic effects demands accurate calibration. Consequently, the accuracy of the radiometer calibration is crucial for converting the measured amplified signal~\cite{wyatt2012radiometric} back to the true emission from the source captured by the antenna.

A particular application of radiometers in astronomy is in the measurement of relative (chromatic or spectral) distortions and of the absolute value of the temperature spectrum from a given celestial source. A classical example of this is the measurement of the Cosmic Microwave Background~\cite{durrer2020cosmic}. The aforementioned calibration challenges are particularly significant when aiming to detect the extremely faint 21-cm cosmological signal from atomic hydrogen at high redshift. This signal is expected to be a few tens or hundreds of milli-Kelvin in amplitude, which is 4--5 orders of magnitude dimmer than the foregrounds that obscure the signal~\cite{pritchard2010constraining}. Radiometers also have multiple uses beyond astronomy and science, being the integral part of many radio receiving systems in fields such as communications~\cite{radar_radiometer}, medical instrumentation~\cite{medical_radiometer} and radar technology~\cite{radar_radiometer}.

Receiver calibration methods can be subdivided into two primary categories. First, there are methods where the calibration ``known'' source is external to the antenna, such as a radio galaxy of known brightness. Second, there are methods where the calibration source is internal and typically connected to the input of the receiver, such as a load resistance. Although the internal approach is less optimal, it is often the only possibility in many applications. For example, when a large field of view prevents dedicated observation of a single external source.

One can also classify receiver calibration methods based on whether they aim to eliminate the effects of the receiver blindly or by fitting a model of the receiver using measured data. Traditional calibration methods used in radio astronomy include techniques such as ``Dicke-switching''~\cite{dicke1946measurement}. In Dicke-switching, the receiver input is continuously switched between the antenna and a reference noise source of constant noise power, which helps mitigate gain fluctuations and receiver noise. Another approach involves using a correlator to reduce the impact of uncorrelated random and systematic noise powers. A well-known alternative is the method known as noise or tone injection~\cite{corbella2000analysis}, where a strong calibration signal is added to the main sky signal without the need to switch the receiver from the antenna to a calibration source. For noise or tone injection to work effectively, the injected calibration signal must be of sufficiently high power relative to the ambient noise and the weak astronomical emissions you are trying to measure. If the astronomical signals are extremely faint, the calibration signal may need to be set at a level that is unrealistically high compared to the noise floor. Such high power levels can be impractical to implement and may even interfere with the measurement process. For example, by saturating the receiver or distorting the system's response.

Recent advances in machine learning offer an attractive alternative for radiometer calibration. In this article, we introduce a technique for both absolute and spectral calibration that uses machine learning to predict the parameters of a receiver model. By connecting a radio receiver to a set of sources with measured physical temperatures and reflection coefficients, and by measuring the corresponding output power spectral densities, we train a neural network to accurately infer the calibration parameters. This method shows significant promise for radio astronomy, providing advantages over traditional analytical techniques. We demonstrate its effectiveness using both measured and simulated data from REACH, a 21-cm sky-averaged cosmology experiment~\cite{de2019reach}.

\subsection{Machine learning for radiometer calibration}\label{sec:MLCal}
\begin{figure}[h]
	\centering
	\includegraphics[width=0.8\textwidth]{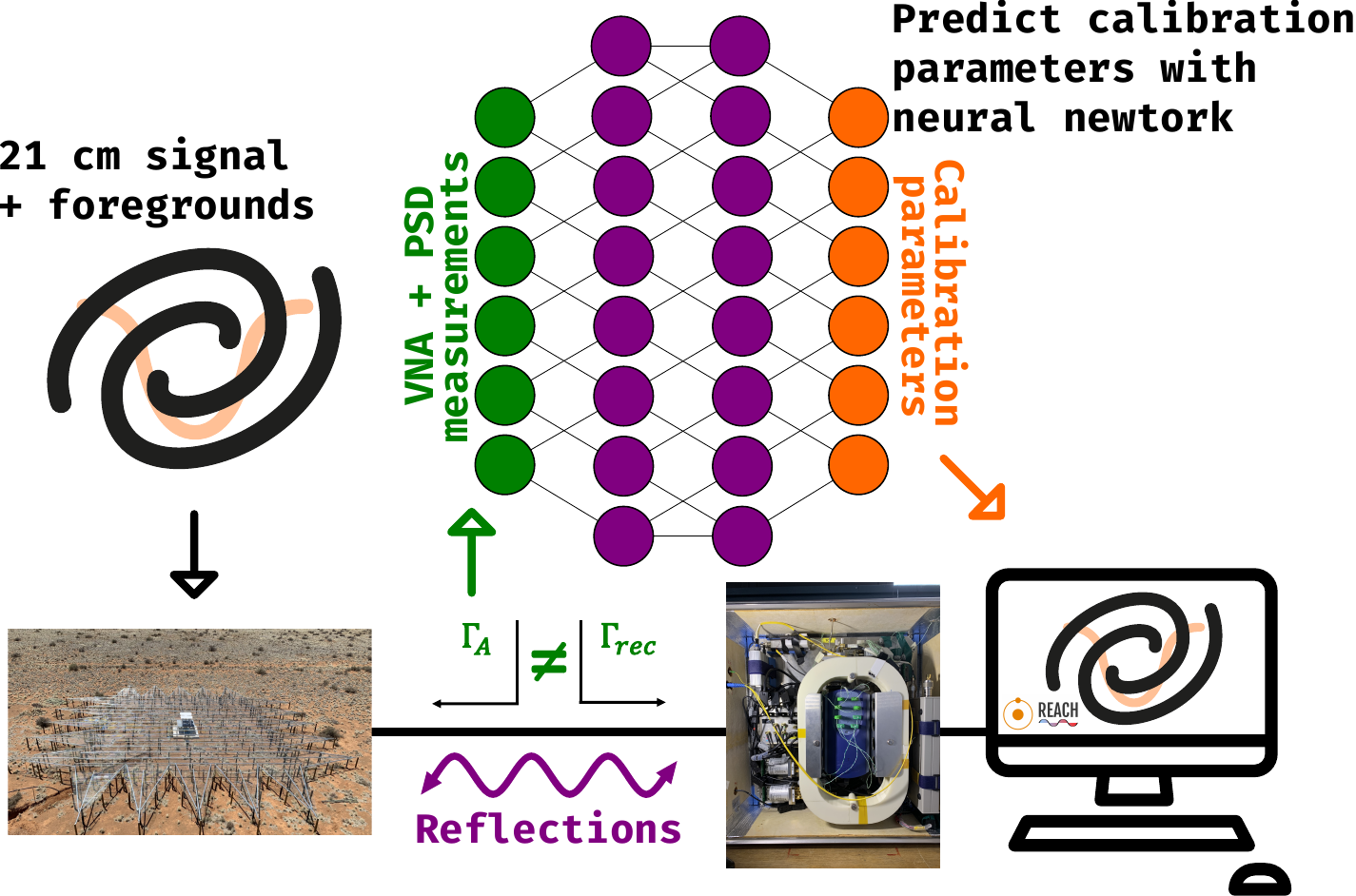}
	\caption{High-level overview of the machine learning-based calibration framework, applied to a 21-cm cosmology experiment. The 21-cm signal obscured by foregrounds~\cite{pritchard201221} is received by the radio telescope as $T_{\text{sky}}$. The signal then passes through an amplifier and other components in the receiver chain. The effect of the receiver chain on the signal is ``calibrated'' by the neural network, which recovers the antenna signal where the 21-cm signal can be extracted. A detailed description of the calibration parameters is provided in Section~\ref{sec:radiometrcalibration}.}
	\label{fig:system}
\end{figure}

We introduce a machine learning-based calibration methodology to model the complex, non-linear behaviours inherent in radiometer systems. A high-level overview of this approach is shown in Figure~\ref{fig:system}, and its advantages compared to traditional methods are summarised in Table~\ref{tab:why_ml}. Neural networks are particularly adept at capturing higher-order effects, making them suitable for calibrating instruments where analytical descriptions are insufficient. Whilst we acknowledge that fitting a derived analytic model of the system parameters would be preferable, in the absence of a complete model, machine learning offers a practical and effective alternative.

A similar calibration methodology was proposed in~\cite{ogut2019deep} and~\cite{alam2023microwave}. However, these methods were designed for systems observing point sources, which can be calibrated by exposing the antenna to previously characterised sources. This is often not possible, for example in sky-averaged experiments where antennas have wide beams and are non steerable. Therefore, new methods are required.

By training the neural network on a diverse set of internal reference sources that encompass the impedance properties of the antenna, we can accurately predict the calibration parameters.

\begin{table}[h]
\centering
\caption{Key advantages of the machine learning-based calibration framework for radiometry, compared against traditional methods. Each advantage is discussed in detail within the indicated sections of this document and supporting references.}
\label{tab:why_ml}
\begin{tabular}{p{0.35\textwidth}p{0.60\textwidth}}
\toprule
\textbf{Advantage} & \textbf{Explanation and Relevant Sections}\\
\midrule

\textbf{No impedance mismatch assumptions} 
& 
Machine learning directly accounts for and learns from measured reflections without requiring ideal matching (see Supplementary Information Section~\ref{sec:traditional_calibration} for a discussion of mismatch assumptions in classical noise wave approaches \cite{rogers2012absolute, monsalve2017calibration}). This avoids systematic offsets that can arise if these assumptions are violated.\\

\textbf{Three-dimensional calibration in frequency, temperature and time domains}
&
The model can predict noise parameter \textit{surfaces}, adding in the temporal axis and thus can naturally handle daily drifts and environmental changes (demonstrated in Supplementary Information B, Section~\ref{sec:temporal}). It is thus well-suited for next-generation space-based 21-cm missions \cite{yamasaki20241, artuc2024spectrometer, bale2023lusee}, where hardware and environmental conditions can shift significantly, and analytical calibration assumptions are often insufficient.\\

\textbf{Frequency-by-frequency calibration} 
& 
The method operates on individual frequency channels rather than relying on frequency-polynomial fits employed in some other commonly used methods~\cite{roque2021bayesian, murray2022bayesian}, preserving finer spectral features. See an example of this in Figure~\ref{fig:karoo_temps}.\\

\textbf{Direct gain recovery} 
& 
The method also recovers the true system gain $g$ (see Section~\ref{sec:results}), highlighted in Figure~\ref{fig:karoo_temps}, which is lost in traditional Dicke-switching-based approaches (derived in Supplementary Information A, Section~\ref{sec:traditional_calibration}).\\

\textbf{Provides absolute calibration} 
& 
The machine learning framework infers the full set of receiver Noise Parameters (Section~\ref{sec:radiometrcalibration}) and maps the measured power spectra onto absolute source temperatures. This enables robust offset recovery for ultra-precise applications whilst reducing dependencies on pre-calibrated reference sources. See Figure~\ref{fig:image1}.\\

\bottomrule
\end{tabular}
\end{table}

We implement this method on the receiver used by the REACH 21-cm radio telescope. The neural network trains on calibration data gathered live in-field whilst the instrument physically switches between internal reference sources, which generates a comprehensive training set. Once the network has converged on the calibration parameters, it then runs on test data gathered as the instrument physically switches to the antenna (or any other previously unseen source, such as those used for demonstration purposes in this work) on which the neural network can make predictions. This strategy effectively learns the complex spectral properties of these receiver calibration parameters, such as the Noise Parameters, a description of which is given in Section~\ref{sec:radiometrcalibration}.

\subsection{Sky-averaged 21~cm Cosmology}\label{sec:21cmintro}
Sky-averaged 21-cm cosmologists are attempting to build the world's most sensitive thermometer to measure the temperature of the early Universe. The redshifted 21-cm line of neutral hydrogen is a faint radio signal that carries the imprint of the first stars and galaxies formed during the Cosmic Dawn and Epoch of Reionisation~\cite{pritchard201221}. By mapping these signals from redshift \textit{z} $\approx$ 6 to 300 (corresponding to observation frequencies on Earth from 200~MHz down to $\sim$ 5~MHz), it is possible to probe the Universe's history from shortly after the Big Bang to the emergence of complex structures, enabling inference on fundamental physics and cosmology~\cite{furlanetto2006cosmology, barkana2018possible}.

In 2018, the EDGES collaboration announced the detection of an absorption profile centred at 78~MHz~\cite{bowman2018absorption}, which was nearly twice as deep as the theoretical models predicted~\cite{furlanetto2006cosmology, barkana2018possible}. This finding was particularly interesting as it suggested potential inaccuracies in the current physical models of the early universe. However, this detection has faced scepticism within the community~\cite{hills2018concerns,sims2020testing,cang2024edges} and was challenged by observations from the SARAS3 telescope~\cite{singh2018saras, singh2018saras2, girish2020saras, singh2022detection}. One of the foremost challenges in observational cosmology today is to verify this measurement or to identify an alternative, true signal.

Despite significant efforts since the potential SARAS3 non-detection in 2022, there have been no new detection. Neither the SARAS3 non-detection nor the EDGES result has been repeated, even though EDGES, SARAS, and now REACH~\cite{de2019reach} are all commissioning new advanced systems and observing in various locations around the globe. The challenge in observational, sky-averaged 21-cm cosmology lies in the signal's extreme faintness, which is predicted to be five orders of magnitude dimmer than the bright radio emission from our own and other galaxies~\cite{pritchard201221}. Therefore, to achieve a confident detection, the radio antenna and receiver must be modelled and calibrated to an unprecedented level of precision and accuracy.

The precision of sky-averaged experiments critically depends on precise radiometer calibration~\cite{razavi2023receiver}. This sensitivity requirement pushes, and in some cases breaks, the limits of current radiometer calibration technology and methodologies. Sky-averaged radiometer calibration is particularly challenging because the antennas used measure the average sky signal. Consequently, they observe the entire sky at all times and are non-steerable. Resultantly, they cannot specifically target well-characterised external sources for calibration, as is standard procedure for other cosmological and astrophysical probes. Therefore, the calibration must be performed using internal reference sources connected to the input of the receiver.

A key challenge lies in the separation of the antenna signal from instrumental effects caused by impedance mismatches between the antenna and internal components~\cite{corbella2002board}. These mismatches lead to reflections that cause standing waves between the antenna and the LNA~\cite{pozar2000microwave}. This results in an interfering signal with complex spectral properties that obscure the true antenna signal as it passes through the receiver. Traditional calibration methods, such as using impedance-matched references and/or employing Dicke-switching~\cite{dicke1946measurement}, rely on assumptions (detailed in Supplementary Information A, Section~\ref{sec:traditional_calibration}) that may break down at the precision levels required to measure the temperature of the early Universe. These techniques often presume that certain components are perfectly impedance-matched and that the system remains stable over time. Whilst these assumptions are generally valid, they can potentially lead to errors~\cite{razavi2023receiver} in the high precision regime required for sky-averaged 21-cm cosmology. Moreover, physical components like switches, cables, and connectors introduce additional complexities, including extra reflections and standing waves that are challenging to model analytically~\cite{razavi2023receiver}. These factors can impact the radiometer's performance, necessitating a calibration approach that can handle such intricacies without depending on potentially incomplete analytical models. As scientists seek fainter signals in the presence of brighter foregrounds, such as the Dark Ages signal~\cite{mondal2023prospects, pritchard2010constraining, fialkov2024cosmic}, and deploy instruments in more demanding environments such as moon surface and orbiting experiments (CosmoCube~\cite{artuc2024spectrometer}, LuSEE Night~\cite{bale2023lusee}, LCRT~\cite{bandyopadhyay2021conceptual} and others~\cite{yamasaki20241, burns2021lunar}), these calibration challenges are expected to become more pronounced. Other challenges in 21-cm modelling include precise antenna beam modelling~\cite{anstey2022informing, hibbard2023fitting, mittal2024impact,sims2023bayesian}. Furthermore, machine learning methods have been employed in tasks such as forecasting~\cite{gessey2024fully} and signal modelling~\cite{bevins2021globalemu, shen2024flexknot, saxena2024simulation, hibbard2024medea}. Precise calibration, as discussed in this work, is essential to enable these sophisticated modelling and mitigation techniques.

\section{Results}\label{sec:results}
Our evaluation consists of two tests. First, in Section~\ref{sec:rr_res}, we calibrate the REACH receiver on an internal load with similar properties to the antenna. Second, in Section~\ref{sec:rda}, we simulate a full receiver chain (incorporating chromatic antenna beam patterns, realistic foregrounds, and an injected sky-averaged 21-cm signal) to assess the overall performance of the method on a representitive science use case.

\subsection{Calibrating the REACH receiver}\label{sec:rr_res}

We calibrate the REACH receiver~\cite{razavi2023receiver} and use it to measure the noise temperature of a known source across frequency (a 69~$\Omega$ resistor on a 2-meter cable) which has similar impedance properties to the REACH dipole antenna. We then compare this result to a direct thermocouple temperature measurement of the resistor to validate the performance of the calibration. We achieve a root-mean-square error (RMSE) on the relative residuals of $0.05~K$ at $1\,\text{MHz}$ channel width within the frequency band of $60$--$130\,\text{MHz}$. Relative residuals are defined as the RMSE error between the predicted and measured temperature values, with the absolute offset accounted for. Calibrating to this level of accuracy and precision on a real antenna would be sufficient to verify or rule out the EDGES signal (mentioned in Section~\ref{sec:intro}) and to detect most possible 21-cm signal models~\cite{bevins2024joint}. This would require the subsequent step of calibrating the chromatic effects introduced by the antenna beam, which is independent of the receiver and is therefore beyond the scope of this paper. The methodology predicts the Noise Parameters on individual frequency channels, meaning that it is sensitive to detailed spectral properties at the resolution of the channel width. Some methods commonly used in the field involve fitting polynomials~\cite{roque2021bayesian, monsalve2017calibration, murray2022bayesian} to the model parameters, which can smooth out and obscure fine spectral features.

\begin{figure}
	\centering
	\includegraphics[width=1\textwidth] {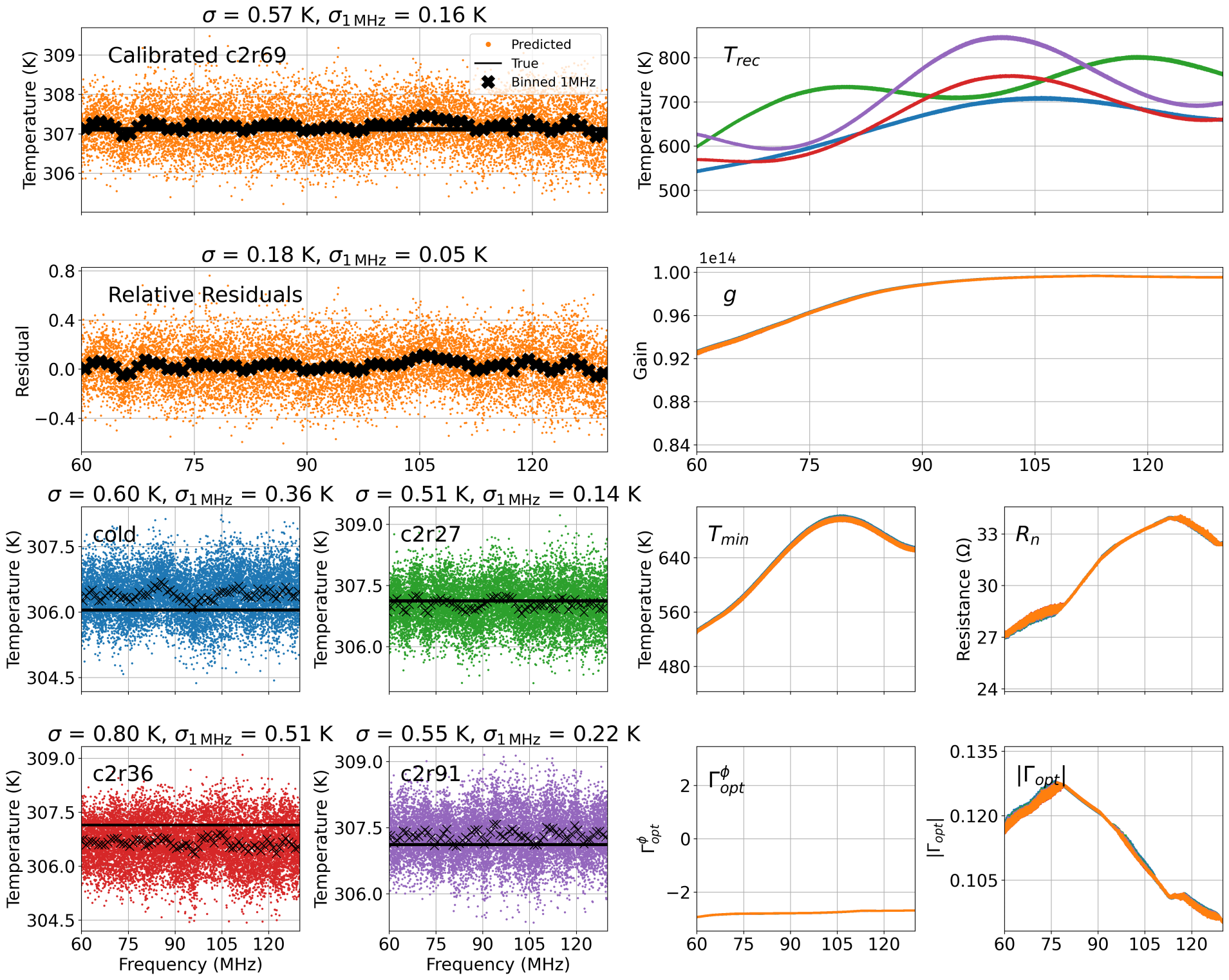}
	\caption{Temperature calibration using internal sources on the REACH receiver. The four panels in the bottom-left quadrant show the measured resistor temperatures (black lines) and the calibrated training solutions (coloured dots). The blue, green, red, and purple dots represent the training data. RMSE errors are shown in the titles. The four panels in the bottom-right quadrant show the predicted Noise Parameters for each source. The top-right two panels show the predicted gain $g$ (source-independent) and receiver temperature $T_{\text{rec}}$ (source-dependent) for each source, calculated from the predicted Noise Parameters. The two panels in the top-left quadrant show the predictions for the unseen source (a $69\,\Omega$ resistor on a 2-meter cable). The ``relative residuals'' represent the error when the data are centred on zero along the temperature axis. All other plots are ``absolute'' (i.e., no offset correction applied). The colours of the predicted resistor temperatures on the left correspond to the plots on the right.}
	\label{fig:karoo_temps}
\end{figure}

Once neural network is trained using the default parameters defined in Section~\ref{sec:hpo}, we show that the REACH receiver can be calibrated to the precision required for subsequent antenna calibration and 21-cm signal detection. Figure~\ref{fig:karoo_temps} shows the calibration results for the REACH receiver obtained in the laboratory from a single observation. In this observation, two separate measurements (each with a 600-second integration time) were taken for each calibration source, and the neural network is trained on both simultaneously. Once training is complete, the network predicts the receiver parameters for an unseen test source. During training, whilst the receiver is switched to the internal calibration sources, high-precision temperature probes measure the true temperature of each source. These measurements are then compared to the predicted noise temperatures, and the resulting loss function is minimised leading to accurate Noise Parameter predictions.

The four panels in the bottom-left of Figure~\ref{fig:karoo_temps}, display the training data used to train the neural network. In these panels, the flat black lines represent the measured temperatures of the reference sources. This data, together with measurements of the output power and reflection coefficients, forms the complete training set. The coloured points indicate the network's predictions for each source after training. At the end of training, the calibration sources were calibrated down to 0.2~K, 0.28~K, 0.67~K, and 0.112~K for the cold, c2r27, c2r36, and c2r91 resistors, respectively (where c(n) represents cable length in meters and r(n) represents load resistance in Ohms); note that the "cold" load corresponds to a c0r50 source.

The panel in the top-left of Figure~\ref{fig:karoo_temps} shows the calibration results for an unseen source (in this case, the c2r69 resistor). Here, the orange points represent the temperature predictions made by the neural network on this test data. The RMSE of the residual error on the normalized noise temperature is 0.05~K when evaluated over 1-MHz channels (represented by the black crosses) within the frequency band of 60--130~MHz. For reference, plausible 21-cm signals are expected to exceed 100~mK in amplitude, and may reach up to 500~mK (the approximate amplitude of the EDGES detection). An absolute calibration RMSE of 0.16~K would be sufficient to either confirm or rule out the EDGES detection, place strong constraints on the presence of an excress radio background and to detect many theoretical 21-cm models.

\subsection{Full chain simulation demonstrating sky-averaged 21-cm signal recovery}\label{sec:rda}

To further evaluate the method's effectiveness once the receiver is connected to the REACH antenna, we simulate a full receiver chain, chromatic antenna beam pattern, realistic foregrounds, and inject a sky-averaged 21-cm signal with properties motivated by current best 21-cm models and constraints to assess whether the method yields residuals sufficiently low to recover a global 21-cm signal under realistic observational conditions. Further details on the simulation methodology are given in Section~\ref{sec:methods}. We show that this signal can be recovered from a realistic simulated system using our machine learning-based framework alongside other current state-of-the-art models in the field.

\begin{figure}
	\centering
	\includegraphics[width=\textwidth]{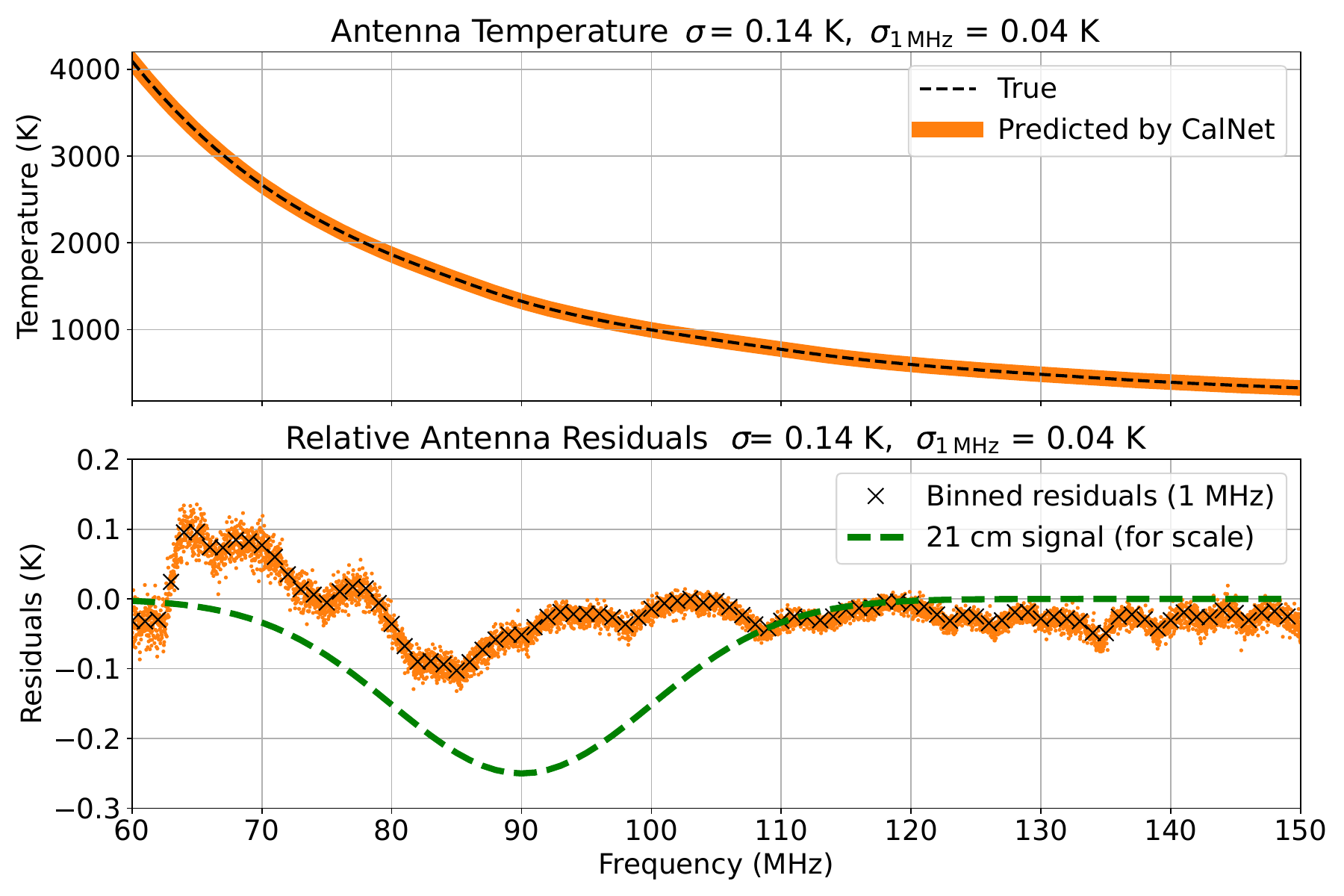}
	\caption{The top panel shows the predicted antenna temperature in orange, with the true temperature overlaid in black dashes. The true signal is the simulated signal, including the foregrounds, beam effects, and simulated 21-cm signal. The predicted temperature is that recovered by the machine learning calibration methodology after the simulated temperatures have passed through the receiver simulation pipeline. The bottom panel shows the relative calibration residuals when the predicted temperatures are subtracted from the true temperatures. The sky-averaged 21-cm signal injected into the simulated antenna temperature is shown in green for scale. The black crosses represent the orange points (originally at a resolution of 12~kHz) binned to 1-MHz channels. $\sigma$ and $\sigma_\text{1~MHz}$ are the RMSE errors when the orange points are subtracted from the true temperature at 12 kHz and 1 MHz, respectively.}
	\label{fig:image1}
\end{figure}

\begin{figure}
    \centering
    \begin{subfigure}[t]{0.545\textwidth}
        \centering
        \includegraphics[width=\textwidth]{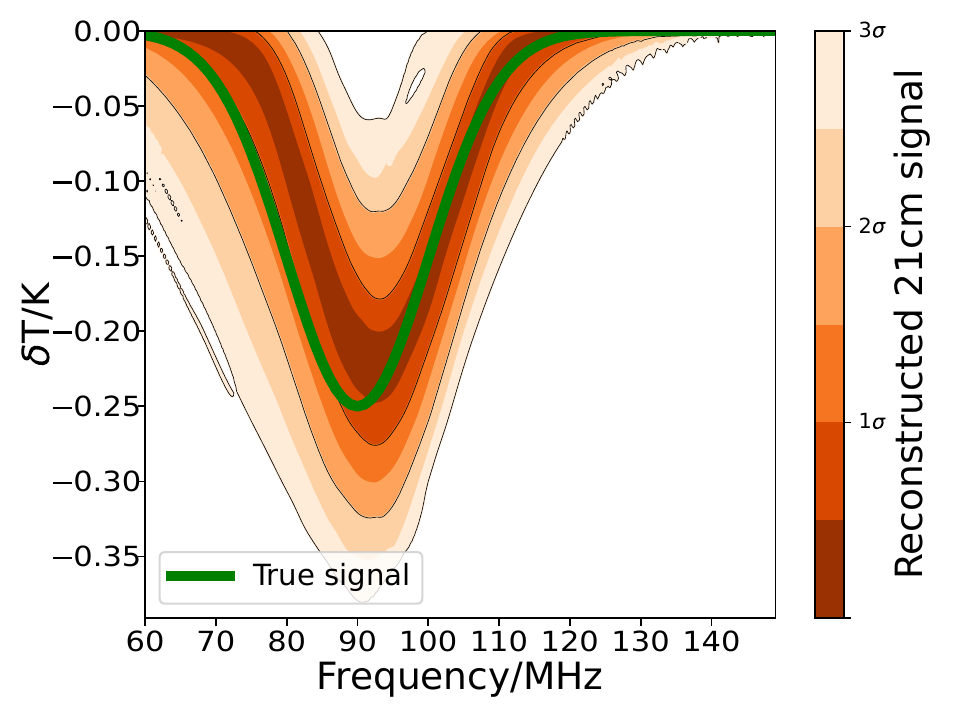}
        \label{fig:image2}
    \end{subfigure}
    \hfill
    \begin{subfigure}[t]{0.435\textwidth}
        \centering
        \includegraphics[width=\textwidth]{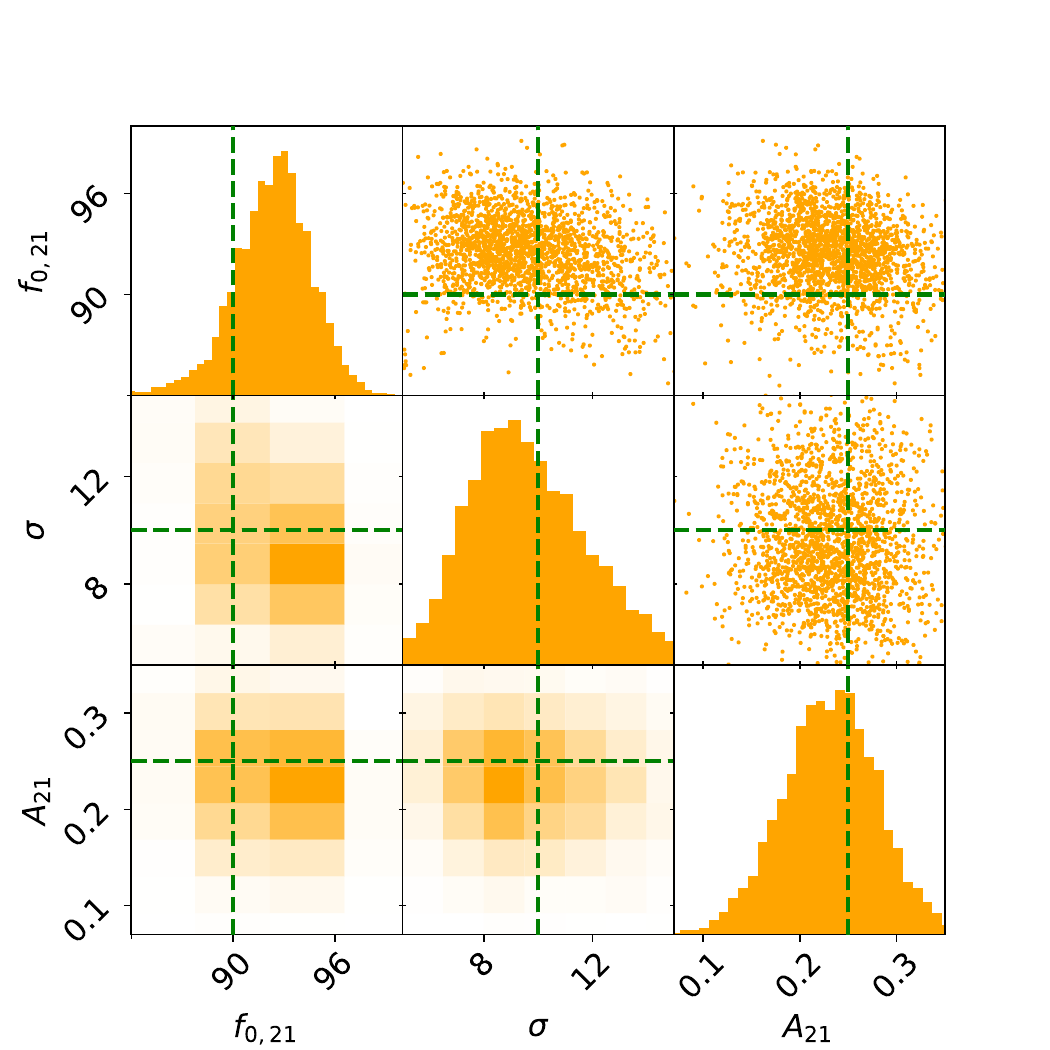}
        \label{fig:image3}
    \end{subfigure}
    \caption{Inferred sky-averaged 21-cm signal from the calibrated antenna temperature shown in Figure~\ref{fig:image1}. Left: The reconstructed sky-averaged 21-cm signal, showing the inferred signal's functional posterior in orange and the initially injected true signal in green. Right: Posterior distributions for the parameters of the recovered sky-averaged 21-cm signal, with the true values indicated in green.}
    \label{fig:combined}
\end{figure}

The simulated antenna is calibrated to an RMSE of 0.14~K at a 12-kHz channel width when observed for 100 hours by the simulated receiver chain. When averaged into 1-MHz resolution channels, the signal is recovered with an RMSE of 0.04~K. We bin down to 1~MHz because it allows us to gain sensitivity whilst ensuring that our channel width is much narrower than the spectral features expected from 21-cm signal models. Although the observation time is conservatively large, a shorter observation period would still suffice to recover a 21-cm signal of this magnitude.

Figure~\ref{fig:image1} shows that calibrating to this residual noise level should, in principle, be sufficient to extract a sky-averaged 21cm signal. Notably, the absolute calibration error and the relative calibration error are the same, indicating that the calibration successfully recovered both the spectral structure and the offset. In contrast, traditional relative calibration methods recover only the spectral structure and require an additional correction for the offset at a later stage.

The calibrated antenna temperature is then passed through the REACH data analysis pipeline, which models the chromatic antenna beam pattern and foregrounds using the methods described in~\cite{anstey2021general, anstey2022informing}. The functional posterior recovered is shown in Figure~\ref{fig:combined} and the distributions over the predicted model parameters are shown in Figure~\ref{fig:combined}. We use the REACH data analysis pipeline for this simulation because it is an advanced forward model that includes beam, chromaticity, horizon, moon and ionosphere modelling ~\cite{pattison2024modelling, anstey2021general, shen2024bayesian}, and Bayesian anomaly mitigation~\cite{leeney2023bayesian, anstey2024enhanced}. Using this, the injected sky-averaged 21-cm signal parameters are recovered as shown in Figure~\ref{fig:combined}.

\section{Discussion}\label{sec:discussion}
In this work, we introduce a novel machine learning-based calibration framework for precision radiometry. Furthermore, we demonstrate its capabilities for application in 21-cm radio cosmology experiments aiming to measure the temperature of the early Universe through the detection of the sky-averaged 21-cm signal. We train a neural network to model the complex, non-linear system behaviour of a precision radiometer. This method, an alternative to traditional calibration approaches, addresses some of the challenges faced by those methods, such as reliance on ideal impedance matching assumptions, limited ability to capture non-linear system behaviour, and difficulties in simultaneously recovering both spectral structure and absolute offsets.

We apply the method to data from the receiver of the REACH radio telescope, and we successfully calibrate most instrumental systematics and reduce the instrumental noise to the noise floor on an internal source chosen for its impedance properties similar to the antenna. On this system, we achieve a relative RMSE of 0.05 K when the predicted temperatures are compared to the measured source temperature. This level of precision, if achieved on the antenna, would be sufficient to verify or rule out the EDGES signal (amplitude $0.5^{+0.42}_{-0.18}$~K~\cite{bowman2018absorption}) and to detect most theoretical models~\cite{bevins2024joint} of the sky-averaged 21-cm signal.

Notably, this methodology operates on a frequency-by-frequency basis, making it sensitive to detailed spectral properties that are sometimes missed by methods relying on fitted polynomials~\cite{monsalve2017calibration, roque2021bayesian}. Furthermore, the method is effective on live, in-situ calibration data and does not require any laboratory measurements. This is critical for highly precise radiometry, where system drifts (demonstrated in Supplementary Information B, Section~\ref{sec:temporal}) could make laboratory measurements taken prior to instrument deployment unreliable. The method also predicts receiver parameters directly, rather than using methods such as Dicke-switching~\cite{dicke1946measurement}, which can lose important gain information and rely on impedance matching assumptions (outlined in Supplementary Information A, Section~\ref{sec:traditional_calibration}) that may not hold at the precision and sensitivity levels required for sky-averaged 21-cm cosmology. It should also be noted that the REACH receiver used in this experiment is more complex than just the LNA, incorporating a second stage of amplification, filtering, an RF-to-optical link, and a high-precision spectrometer. In practice, REACH plans to combine this type of machine learning calibration with other more traditional methods, capable of providing better insight into the physics behind the calibration solution.

Our initial efforts focus on calibrating the REACH receiver using internal reference sources, to ensure the system is ready for precise and accurate antenna measurements upon completion of commissioning. Through extensive and highly detailed simulations incorporating realistic models of the antenna~\cite{cumner2022radio}, receiver chain~\cite{sun2024calibration}, chromatic beam patterns~\cite{anstey2022informing, cumner2024effects}, and foregrounds~\cite{anstey2023enhanced}, we have also demonstrated that our machine learning-based calibration framework can calibrate receiver instrumental systematics sufficiently well that the sky-averaged 21-cm signal could be recovered from a simulated realistic antenna. These simulations include an injected sky-averaged 21-cm signal model based on current best 21-cm models and constraints~\cite{bevins2024joint}.

Furthermore, the machine learning-based calibration framework not only addresses current challenges in radiometer calibration for sky-averaged 21-cm cosmology but also opens new avenues for future research and instrumentation. By accurately modelling complex, non-linear system behaviours without relying on simplifying assumptions, this approach can be adapted to other instruments and observational setups. This adaptability is particularly significant as the field moves toward more ambitious projects, such as deploying radiometers in space~\cite{yamasaki20241, artuc2024spectrometer, bale2023lusee} or on the lunar surface~\cite{reising2016tropospheric, bandyopadhyay2021conceptual, burns2021lunar}. There may also be applications beyond cosmology, such as airborne radiometry (where alternative machine learning-based methods have been used for radiometer calibration~\cite{bosch2019instrument, alam2023microwave}) and other areas of astrophysics requiring highly sensitive measurements. Future work will focus on incorporating larger and more diverse datasets, particularly large sets of observations spanning many nights that can be fitted simultaneously.

\section{Methods}\label{sec:methods}
In this section, we explain our machine learning-based radiometer calibration methodology in detail.

\subsection{Radiometer calibration}\label{sec:radiometrcalibration}
Radiometer calibration is the process of mapping measured power spectra onto input source temperatures. The goal of ``calibration'' is thus to build a model that links the input temperature $T^{\text{in}}_{\text{src}}$ to the output power $\mathcal{P}_{\text{src}}^{\text{out}}$. Typically, radiometers are calibrated by exposing the antenna to a known source, such as an internal black-body~\cite{kendall1970two, sapritsky1995black} or a well-characterised external source~\cite{adam2016planck, noordam2004lofar}. However, the latter is not possible for sky-averaged experiments as the antenna beam covers the whole sky and is not directive.


\begin{figure}
	\centering
	\includegraphics[width=\textwidth]{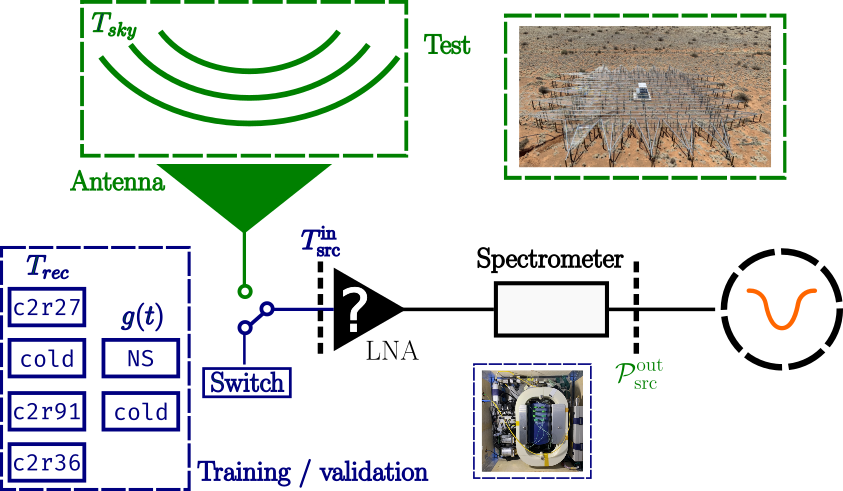}
	\caption{A physical system is built to generate data that represents the impedance properties of the antenna. The system generates training data from the calibration sources then switches to the antenna to generate test data where predictions can be made. In Section~\ref{sec:rr_res}, a characterised internal source is used in place of the antenna and in Section~\ref{sec:rda} the antenna is simulated.}
	\label{fig:instrument}
\end{figure}
For sky-averaged experiments, the calibration methodology proposed by Dicke~\cite{dicke1946measurement} is commonly used to calibrate the non-linear, time-dependent gain of the LNA. This technique involves comparing the source temperature ($T^{\text{in}}_{\text{src}}$) with internal reference loads whose properties are known: a noise source ($T^{\text{in}}_{\text{NS}}$) and a load ($T^{\text{in}}_{\text{L}}$). We use resistors because they emit well-understood Johnson-Nyquist noise, which is proportional to physical temperature, so their true power output can be inferred from a temperature measurement. It should be noted that these methods either assume that the reference loads are impedance-matched to each other, which is difficult to achieve in reality, or apply a first-order correction~\cite{roque2021bayesian, monsalve2017calibration, murray2022bayesian}. Although this assumption and subsequent first-order correction are generally reasonable, they can potentially lead to non-negligible errors (see Supplementary Information A, Section~\ref{sec:traditional_calibration}) at the fine scales required for sky-averaged 21-cm cosmology. Other methods assume ``ideal'' open and short cables~\cite{price2018design, price2022measuring}, which, although also generally reasonable, may lead to errors in this high-precision regime.

In addition to the non-linear gain $g$ introduced by the amplifier, an impedance mismatch between the source of interest and the LNA modulates the amplifier’s intrinsic noise, resulting in standing waves along the cable that contribute an additional temperature component, $T_{\text{rec}}$, to the output signal. Therefore, as shown in Equation~\ref{eq:base_t}, $T^{\text{in}}_{\text{src}}$ can be related to $\mathcal{P}_{\text{src}}^{\text{out}}(\nu)$, where we also define the mismatch factor:

\begin{equation}\label{M} M = \frac{(1 - |\Gamma_{\text{src}}|^2)(1 - |\Gamma_{\text{rec}}|^2)}{|1 - \Gamma_{\text{src}} \Gamma_{\text{rec}}|^2}, \end{equation} which preserves the source independence of the gain $g$~\cite{pozar2000microwave}.

The Noise Parameter formalism introduced by~\cite{haus1960representation} accounts for the effect of these standing waves, so that Equation~\ref{eq:base_t} becomes:

\begin{equation}\label{eq:noise_g} \mathcal{P}{\text{src}}^{out} = gM \bigg( T^{in}{\text{src}} + T_{min}+T_0\frac{4R_N}{Z_0}\frac{|\Gamma_{\text{src}}-\Gamma_{\text{opt}}|^2}{(1-|\Gamma_{\text{src}}|^2)|1+\Gamma_{\text{opt}}|^2} \bigg), \end{equation}

where $\Gamma_{\text{rec}}$ and $\Gamma_{\text{src}}$ are the measurable reflection coefficients of the receiver and source, respectively, as measured by a Vector Network Analyser (VNA), and $Z_0$ and $T_{0}$ are measurable properties of the LNA. It is important to note that these two values are constants and are degenerate with $R_N$. Whilst their true values are measured to ensure unit consistency, they could be set to 1, and the fit would still yield equivalent results. The remaining parameters—$|\Gamma_{\text{opt}}|$, $\Gamma_{\text{opt}}^{\phi}$, $R_N$, and $T_{min}$—must be predicted to characterize $T_{\text{rec}}$. Uncalibrated values of $\mathcal{P}_{\text{src}}^{out}$ are measured by a spectrometer, which determines the intensity of electromagnetic radiation at various wavelengths.

The Noise Parameter formalism is proposed in the context of sky-averaged 21-cm cosmology in~\cite{price2022measuring}. An alternative approach, used more commonly in this context, represents the system using Noise Wave parameters, as introduced in~\cite{meys1978wave}, more recently applied to astronomy in~\cite{rogers2012absolute}, and applied directly to sky-averaged 21-cm cosmology in~\cite{monsalve2017calibration, bowman2018absorption, roque2021bayesian}.
The Noise Wave and Noise Parameter formalisms are mathematically equivalent and both (assuming matching assumptions are correct or lead to negligible error) leave four unknown source independent parameters to be computed. For a more detailed explanation see Supplementary Information A, Section~\ref{sec:traditional_calibration} and we recommend the aforementioned literature.

\subsection{Machine learning}\label{sec:machinelearning}
A neural network is composed of a series of interconnected layers of nodes. Nodes represent mathematical functions called activation functions that act on signals passing through them. Nodes are connected by weights that vary the amplitude of the signal as it passes through. Given specific inputs (data), the weights can be adjusted so that the outputs become more desirable relative to the inputs by differentiating a goodness-of-fit measure, known as the ``loss,'' with respect to the weights. By adjusting the weights to minimise the loss, neural networks can make accurate predictions based on complex non-linear relationships between inputs and outputs. A neural network as described above is known as a multi-layer perceptron.

At its core, radiometer calibration requires mapping output powers $\mathcal{P}_{\text{ant}}^{\text{out}}$ onto input temperatures of interest, $T^{\text{in}}_{\text{ant}}$. We characterise this relationship using a neural network trained to predict Noise Parameters based on thermocouple data, power spectral density (PSD) measurements, and Vector Network Analyser (VNA) measurements. Machine learning datasets are typically divided into three sets: \textit{training}, \textit{validation} and \textit{test}. The neural network regresses over the training data, cross-validates on the validation data, and the test data remains unseen to provide an unbiased assessment. These sets are normally generated by simply splitting a broader dataset into three according to some ratio.

\subsubsection{Training data}
The neural network trains on carefully designed and selected internal sources and then tests on data collected when the system physically switches to the target source, allowing it to make predictions, as shown in Figure~\ref{fig:instrument}. Since it is not possible to build a single source representing the antenna's complex impedance, the system switches between a series of electrical components, each with different impedance properties that sample the overall region of the Smith chart occupied by the antenna~\cite{razavi2023receiver}. A Smith chart is a graphical tool used to represent complex impedance and reflection coefficients.
If the calibration sources are representative of the antenna, the neural network should be able to learn how to predict the Noise Parameters for the calibrators and subsequently predict them for the antenna.

\subsubsection{Architecture}\label{sec:arch}
We aim to train a neural network to characterise the receiver such that, based on thermocouple data and VNA measurements, the true antenna temperature can be accurately predicted on a frequency-by-frequency basis. This was done in~\cite{ogut2019deep} by training on and predicting antenna temperatures directly. However, this is not possible using current sky-averaged 21-cm cosmology instrumentation, as the sky temperature reaches up to 5000 K in the frequency band of interest, and current instruments are not built with sources (or training data) at these temperatures. This is due to the practical constraints of heating a resistor to 5000 K. Furthermore, unlike in~\cite{ogut2019deep2}, sky-averaged antennas are not steerable and cannot target known sources for training. Neural networks are built to interpolate, not extrapolate. Therefore, we must design our architecture such that the network predicts and trains on data that are informative and cover a similar range for both the training data (the calibration sources) and the test data (the antenna). Conveniently, the source-independent Noise Parameter formalism described in Section~\ref{sec:radiometrcalibration} provides us with this.

Therefore, the neural network must be designed to predict the Noise Parameters given thermocouple data, PSD data, and VNA measurements. The Noise Parameters predicted by the neural network can then be used in a single analytical step to predict $T^{\text{in}}_{\text{src}}$. We use a multi-layer perceptron, with the number of layers and neurons to be determined by a hyper-parameter optimisation methodology described in Section~\ref{sec:hpo}.

\begin{figure}
	\centering
	\includegraphics[width=0.8\textwidth]{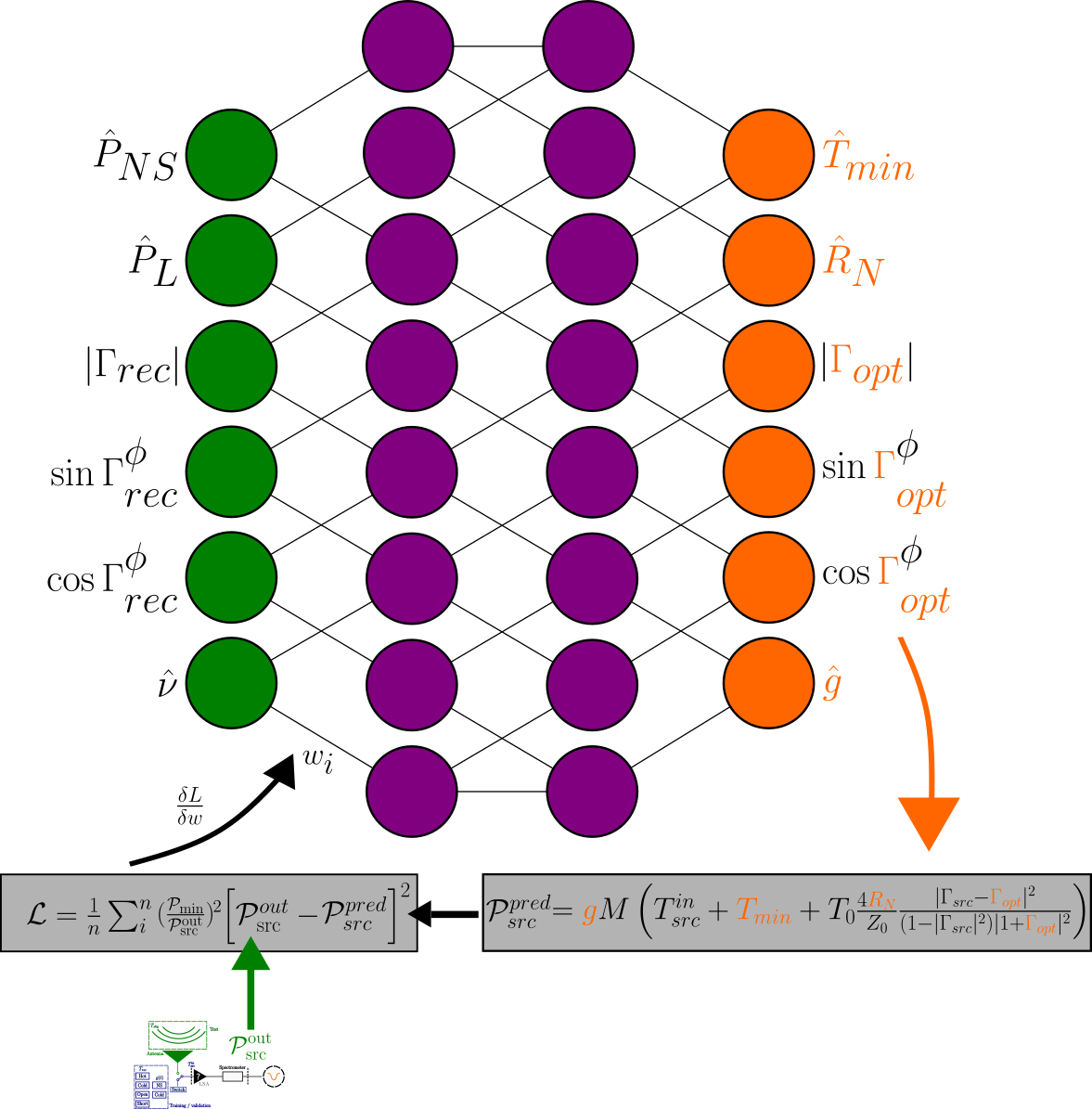}
	\caption{Depicts network architectures, showing the flow from predicted Noise Parameters, to output power calculation, to loss, to weights optimisation. The inputs from the PSD measurements and VNA measurements are shown in green and the outputs in orange. See Figure~\ref{fig:instrument} for an enlarged view of the instrument, which outputs $\mathcal{P}_{src}^{out}$. Values with a  are normalised.}
	\label{fig:nn}
\end{figure}

\subsubsection{Loss function}
Given the network architecture described in Section~\ref{sec:arch}, a loss function must be defined that quantifies the ``goodness-of-fit'' of the predicted Noise Parameters. The true values of the Noise Parameters are not known so cannot be regressed over directly. However, the true temperature of the calibration sources can be measured. Equation~\ref{eq:noise_g} links output power to input temperature given Noise Parameters and other measurable properties. As such, we can predict the Noise Parameters and use them alongside other measured parameters to regress over the mean squared error between the measured power (acting as ground truth) and the predicted power. The predicted power is calculated using the neural network to predict the Noise Parameters, followed by an analytical step using Equation~\ref{eq:noise_g} to compute $\mathcal{P}_{\text{src}}^{\text{out}}$. This yields the loss function
\begin{equation}
	\mathcal{L} = \frac{1}{n} \sum^n_i (\frac{\mathcal{P}_{\text{min}}}{\mathcal{P}_{\text{src}}^{out}})^2\Bigl[ \mathcal{P}_{\text{src}}^{out} -
	g M \left( T^{in}_{\text{src}} + T_{\text{min}} + T_0\frac{4R_N}{Z_0}\frac{|\Gamma_{\text{src}}-\Gamma_{\text{opt}}|^2}{(1-|\Gamma_{\text{src}}|^2)|1+\Gamma_{\text{opt}}|^2} \right) \Bigr]^2,
\end{equation}
where the term $(\frac{\mathcal{P}_{\text{min}}}{\mathcal{P}_{\text{src}}^{\text{out}}})^2$ is included to account for the fact that the noise in the system is radiometric, not Gaussian, as is often assumed when using a mean-squared error loss function. $\mathcal{P}_{\text{min}}$ is the lowest measured power across the training set.

Visualised in Figure~\ref{fig:nn}, this loss function allows the estimation of how close the predicted Noise Parameters are to their ``true'' values without explicitly knowing these true values. The gradient of the high-dimensional loss landscape with respect to the network weights can be approximated using automatic differentiation (we use TensorFlow). From this, the loss can be adjusted towards its minima. Thus, the true Noise Parameters can be estimated and subsequently $\mathcal{P}_{\text{src}}^{\text{out}}$ calculated, which can be inverted to yield $T_{\text{src}}^{\text{in}}$, which in the context of cosmology will be $T_{\text{sky}}^{\text{in}}$.

\subsubsection{Data normalisation}
Neural networks work most efficiently and optimally on values of order unity, as this prevents the saturation of activation functions. We normalise the input powers and frequencies to within this range using min-max normalisation~\cite{geron2022hands}. Training on complex numbers is slightly more intricate, as the system must be designed to interpret the wrapped nature of such parameters. $0\pi$ represents the same point in the phase cycle as $2\pi$, but in a physical scenario these two positions in phase may have different practical implications. We can encode this into the data directly using phase wrapping. As shown in Figure~\ref{fig:nn}, we achieve this by feeding the network $|\Gamma_{\text{rec}}|$, $\sin \Gamma_{\text{rec}}^{\phi}$, and $\cos \Gamma_{\text{rec}}^{\phi}$.

The system we aim to predict is physical, but many degenerate non-physical solutions form troughs in the loss landscape. This can slow or prevent convergence on the physical solution. To solve this, we encode physics into our network architecture where possible by constraining $R_{N}$, $g$, and $T_{\text{min}}$ to be positive, scaling them between zero and one based on prior knowledge of their maximum scale, and fixing their output activation to a sigmoid function. $\Gamma_{\text{opt}}$ is complex, and we predict its absolute value and the unwrapped phase as described for the inputs. Although convergence is achieved with or without this scaling, it can be significantly accelerated by constraining network outputs to tighter ranges based on loose prior knowledge of the Noise Parameters.

\subsubsection{Hyper-parameter optimisation}\label{sec:hpo}
The system hyperparameters are Network structure, learning rate, dropout (proportion of nodes randomly ``turned off'' during training, for regularisation), and batch size are optimised using a Bayesian optimisation algorithm. We use Hyperopt~\cite{bergstra2015hyperopt}, a Python library, which implements Bayesian optimisation through the Tree-structured Parzen Estimator (TPE) algorithm~\cite{watanabe2023tree}. This method systematically improves upon candidate solutions based on prior evaluations, utilising a probabilistic model to predict the performance of new hyperparameter combinations. Based on these results, we can select the optimal network structure and hyperparameter combination. Throughout the training process we implement early stopping when the validation loss does not improve for 1000 epochs, with the total number of epochs being set to 5000. An epoch is 1 training loop over the full training set.


Optimal values depend on data size and complexity, so further optimisation may be required for future use. The optimal parameters determined using the REACH receiver data were: batch size of 256, dropout uniformly sampled between 0 and 0.5 on all nodes, learning rate of 0.0005, 60 neurons in layer 1, 100 neurons in layer 2, and 80 neurons in layer 3.

\subsection{Simulated data}
 To validate our machine learning-based calibration framework, we assess extensive simulations mimicking the operational conditions of a sky-averaged 21-cm experiment, as detailed in Section~\ref{sec:rda}. In this section, we explain how the data for these physical simulations were generated. These simulations incorporated realistic models of the REACH antenna, receiver chain, chromatic beam patterns, and foreground emissions. We utilised the methodologies described in~\cite{sun2024calibration} to simulate the receiver chain and those in~\cite{anstey2021general, anstey2022informing} to generate synthetic datasets closely resembling the data expected from the REACH receiver.

The simulated antenna parameters were derived to be similar to the characteristics of the proposed REACH antenna, including impedance properties and beam patterns~\cite{anstey2022informing}. A realistic sky model was incorporated, accounting for foreground emissions from our galaxy and extragalactic sources as outlined in~\cite{anstey2021general}. The receiver chain was modelled using Noise Parameter measurements obtained from laboratory tests of the REACH receiver components~\cite{razavi2023receiver}.

To achieve a comprehensive simulation of the entire calibration process, our workflow proceeded as follows: \begin{enumerate} \item \textbf{Antenna Temperature Simulation:} Beam simulations and realistic foreground models (as detailed in~\cite{anstey2021general, anstey2022informing, sun2024calibration, cumner2022radio}) were used to generate a realistic sky-averaged 21cm antenna temperature. \item \textbf{Receiver Chain Simulation:} The simulated antenna temperature was passed through a comprehensive simulation of the receiver, which incorporates laboratory measurements from the REACH receiver, to produce the corresponding simulated output power. \item \textbf{Reference Load Simulation:} In parallel, output power for simulated reference loads (mimicking the calibration sources) was generated. \item \textbf{Neural Network Training:} The simulated measurements from the reference loads were used to train a neural network to accurately predict the receiver’s Noise Parameters. \item \textbf{Antenna Temperature Recovery:} The trained neural network was subsequently applied to recover the antenna temperature from the simulated receiver output. \item \textbf{Signal Extraction:} A model was fitted to the recovered antenna temperature to extract the underlying sky-averaged 21-cm signal, based on the most current constraints~\cite{bevins2024joint}. \end{enumerate}

A theoretical model of the sky-averaged 21-cm signal was injected into these simulations to assess the capability of our calibration framework to recover the signal amidst instrumental effects, systematics, and environmental factors that might impact the observations.

\subsection{REACH receiver data} For the practical implementation of our calibration framework, we used data from the REACH (Radio Experiment for the Analysis of Cosmic Hydrogen) radiometer. The REACH instrument is designed specifically for precision measurements of the sky-averaged 21-cm signal, featuring a custom-built antenna and receiver system optimised for low-noise observations.

We collected the calibration data by switching the receiver between multiple internal reference sources with known impedance properties, as well as an internal source with properties similar to the antenna and measurable ground truth temperature for evaluation. The system design and Smith chart coverage for this configuration are outlined in~\cite{razavi2023receiver}. The methodology for data acquisition and instrument configuration follows the procedures detailed in~\cite{de2019reach, razavi2023receiver}. The internal reference sources are carefully designed to sample the impedance space occupied by the antenna, thereby providing a comprehensive training set for the neural network.

Temperature measurements from thermocouples attached to the calibration sources were recorded to obtain accurate reference temperatures. Additionally, Vector Network Analyser (VNA) measurements were performed to characterise the reflection coefficients of the receiver and sources. These measurements served as crucial inputs to the neural network during the training phase.

\section*{Contributions}
S.A.K. Leeney led the research, development, and preparation of the manuscript. H.T.J. Bevins supervised the research and proposed the initial concept for the study. E. de Lera Acedo supervised the project and provided guidance on the fundamentals of radiometer calibration. W.J. Handley supervised the project, offering advice on machine learning and data analysis. C. Kirkham advised on simulated data generation and preprocessing for calibration. R. S. Patel configured and ran the observations on the REACH receiver.  J. Zhu developed the initial calibration simulation pipeline. D. Molnar provided foundational guidance on radiometer calibration during the early stages of the project. D. Anstey helped configure the REACH data analysis pipeline for the analysis of simulated data, and J. Cumner contributed to the generation of simulated antenna temperatures. The remaining authors, from K. Artuc to M. Spinelli, contributed to the REACH telescope and are therefore on the REACH builders list. They also contributed to the manuscript's preparation and review. These authors are credited in alphabetical order.

\newpage

\section{Supplementary Information A: Assumptions In Noise Wave Parameter Methods}\label{sec:traditional_calibration}

When calibrating a receiver in the Noise Wave Parameter framework, the reflection coefficients of the calibration source ($\Gamma_{\mathrm{cal}}$) and the receiver itself ($\Gamma_{\mathrm{rec}}$) are measured, along with the power spectral densities (PSDs) of the calibration source ($P_{\mathrm{cal}}$), the internal reference load ($P_{L}$), and the internal noise source ($P_{\mathrm{NS}}$) \cite{monsalve2017calibration,bowman2018absorption}. These measurements allow for a preliminary, uncalibrated temperature calculation, commonly denoted $T^*_{\mathrm{cal}}$, to be computed via Dicke-switching using:
\begin{equation}
T^*_{\mathrm{cal}} \;=\; T_{\mathrm{NS}}
\,\biggl(\,\frac{P_{\mathrm{cal}} \;-\; P_{L}}{P_{\mathrm{NS}} \;-\; P_{L}}\biggr)
\;+\; T_{L},
\end{equation}
where $T_{L}$ is an assumed value for the noise temperature of the internal reference load, and $T_{\mathrm{NS}}$ is the assumed excess noise temperature of the internal noise source. This step removes system gain variations arising from cables, filters, amplifiers, and the analogue-to-digital converter~\cite{monsalve2017calibration}, under the assumption that...

Each PSD measurement may then be written in the form:
\begin{align}
P_{\mathrm{cal}} &\;=\; g_{\mathrm{sys}}
\Bigl[\,
T_{\mathrm{cal}}
\bigl(1 - \bigl|\Gamma_{\mathrm{cal}}\bigr|^{2}\bigr)
\Bigl\lvert
\frac{\sqrt{\,1-\bigl|\Gamma_{\mathrm{rec}}\bigr|^{2}}}
{\,1 - \Gamma_{\mathrm{cal}}\,\Gamma_{\mathrm{rec}}\,}
\Bigr\rvert^{2} \;+\;
T_{\mathrm{unc}}\,
\bigl|\Gamma_{\mathrm{cal}}\bigr|^{2}\,
\Bigl\lvert
\frac{\sqrt{\,1-\bigl|\Gamma_{\mathrm{rec}}\bigr|^{2}}}
{\,1 - \Gamma_{\mathrm{cal}}\,\Gamma_{\mathrm{rec}}\,}
\Bigr\rvert^{2} \notag \\
& \;+\;
T_{\mathrm{cos}}\,
\Re\!\Bigl[\,
\frac{\Gamma_{\mathrm{cal}}}{1-\Gamma_{\mathrm{cal}}\,\Gamma_{\mathrm{rec}}}
\Bigr]
\sqrt{\,1-\bigl|\Gamma_{\mathrm{rec}}\bigr|^{2}}
\;+\;
T_{\mathrm{sin}}\,
\Im\!\Bigl[\,
\frac{\Gamma_{\mathrm{cal}}}{1-\Gamma_{\mathrm{cal}}\,\Gamma_{\mathrm{rec}}}
\Bigr]
\sqrt{\,1-\bigl|\Gamma_{\mathrm{rec}}\bigr|^{2}}
\;+\;
T_{0}
\Bigr],
\end{align}

where $g_{\mathrm{sys}}$ is the system gain referenced to the receiver input, and $T_{\mathrm{cal}}$ denotes the calibrated input temperature. The terms $T_{\mathrm{unc}}$, $T_{\mathrm{cos}}$, and $T_{\mathrm{sin}}$ are the "noise wave parameters" introduced by Meys~\cite{meys1978wave}, which have been employed in global 21-cm experiments~\cite{monsalve2017calibration,rogers2012absolute,bowman2018absorption}. In these methods, the internal load and noise sources are assumed to have negligible reflection coefficients (often approximated as zero when $|\Gamma|\!<\!0.005$). This is illustrated by:
\begin{equation}
P_{L} = g_{\mathrm{sys}}^{*} \Bigl[ T_{L} \bigl(1 - \bigl|\Gamma_{\mathrm{rec}}\bigr|^{2}\bigr) + T_{0}^{*} \Bigr],
\end{equation}

\begin{equation}
P_{\mathrm{NS}} = g_{\mathrm{sys}}^{*} \Bigl[ \bigl(T_{L} + T_{\mathrm{NS}}\bigr) \bigl(1 - \bigl|\Gamma_{\mathrm{rec}}\bigr|^{2}\bigr) + T_{0}^{*} \Bigr].
\end{equation}

When the internal references are located on a different reference plane from that of the receiver input, they introduce offsets $g_{\mathrm{sys}}^{*}$ and $T_{0}^{*}$ distinct from the values in the main signal path \cite{monsalve2017calibration}. EDGES-style approaches account for these using extra scale and offset parameters \cite{monsalve2017calibration,bowman2018absorption}, which can also absorb uncertainties in $T_{L}$ and $T_{\mathrm{NS}}$.

Both the Noise Wave and Noise Parameter methods use two sets of sources: one for Dicke-switching and another for calibration parameter estimation (Noise Wave Parameter or Noise Parameter). The sources used for Dicke-switching must be matched, whereas the sources used for calibration parameter estimation do not need to be matched.

Finally, combining the above expressions produces a linear relationship between the \emph{uncalibrated} temperature and the final, \emph{calibrated} temperature. The reader should note that throughout such treatments, several assumptions are made:
(i) low-reflection internal references,
(ii) negligible mismatch between the internal references, and
(iii) parametric modelling of noise waves via polynomials in frequency ($T_{\mathrm{unc}}, T_{\mathrm{cos}}, \dots$).

In experiments where the required precision presses well beyond the validity regime of these approximations (as is often the case for sky-averaged 21-cm experiments), systematic errors may be introduced. By contrast, our machine learning approach predicts the necessary parameters (specifically, the Noise Parameters) without \emph{a priori} assumptions of perfect impedance matching or negligible reflections in the internal references. Consequently, our framework avoids introducing potential modelling biases inherent in the traditional scheme, providing a more comprehensive solution for high-precision sky-averaged 21-cm radiometry.

\section{Supplementary Information B: Training on a Temporally Unstable System}\label{sec:temporal}

To demonstrate how the method operates on an unstable system, we generated a simulated dataset designed to mimic a temporally unstable system, reflecting the complexities encountered in real-world observational environments. Similar to the process described in Section~\ref{sec:rda} for generating simulated data, we simulated a realistic model of the receiver chain. However, in this instance, we extended the simulation to cover multiple observations, accumulating a total of 1000 observations, each lasting 30 seconds, for every calibration source. This dataset allows us to test our calibration framework in the temporal domain, predicting three-dimensional Noise Parameter \textit{surfaces} (parameter value as a function of frequency and time), rather than just two-dimensional frequency-dependent curves.

\begin{figure}[h!]
    \centering
    \includegraphics[width=\textwidth]{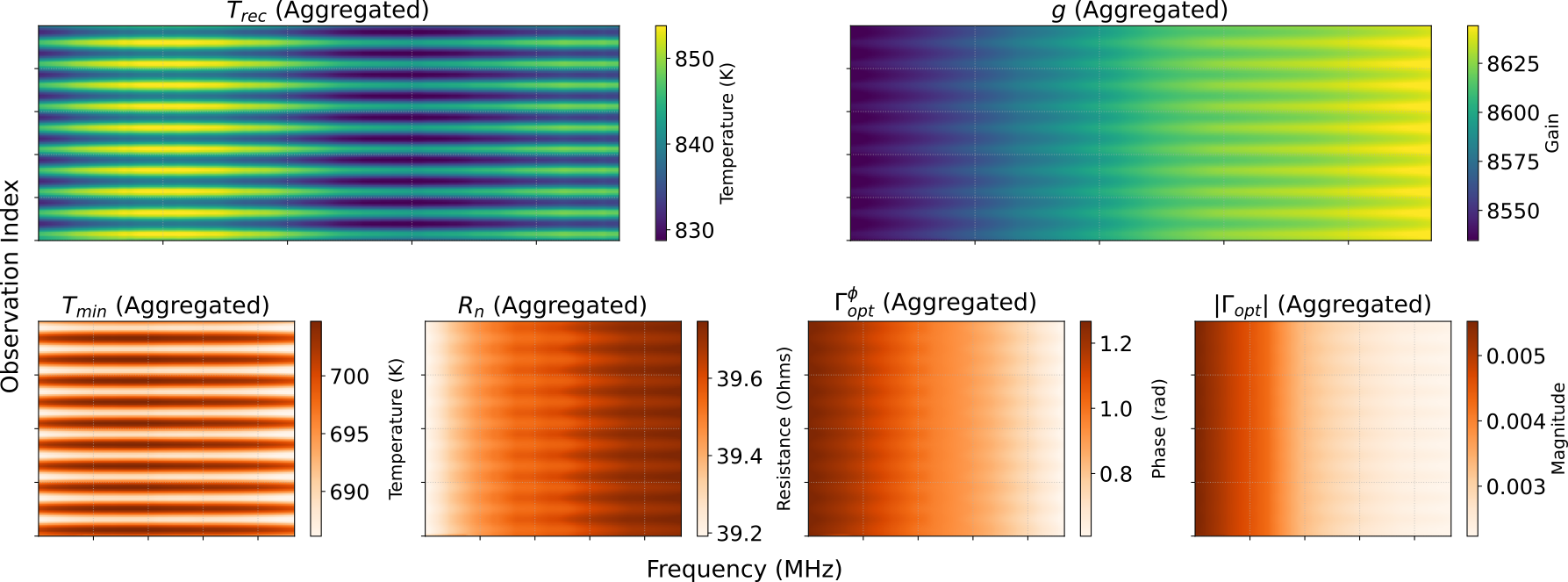}
    \caption{Heatmaps illustrating the temporal evolution of the inferred receiver Noise Parameters over the simulated multi-night observation period. Each panel corresponds to one of the four Noise Parameters ($T_{\text{min}}$, $R_N$, $|\Gamma_{\text{opt}}|$, $\Gamma_{\text{opt}}^{\phi}$) and the gain $g$. The y-axis represents observation order (ascending from first to last), and the x-axis represents frequency. Colour intensity indicates the inferred parameter value. The sinusoidal variation deliberately introduced into $T_{\text{min}}$ (bottom-left panel) is clearly visible and accurately recovered by the neural network, demonstrating the model's ability to characterise a temporally unstable system. Note that each heatmap contains aggregated data from all calibrators used in the simulation.}
    \label{fig:temporal_heatmap}
\end{figure}

A key distinction in this simulation was the introduction of temporal variability into one of the system's Noise Parameters. Specifically, the Noise Parameter $T_{\text{min}}$ was varied sinusoidally by a fixed offset over the course of the observations, fluctuating by $\pm 5\%$ around its nominal value. This variation was implemented to simulate the gradual drifts and changes that can occur in the electronic components of a receiver due to environmental factors or ageing, representing a significant challenge for traditional calibration techniques. It is important to note that many conventional calibration methods rely on the assumption that receiver Noise Parameters remain stable over the observation period, with the exception of the overall gain $g$. Such temporal instabilities, if not properly accounted for, can introduce systematic errors that corrupt the final scientific measurements.

We trained the neural network using this temporally dynamic dataset. The network successfully learned to characterise the unstable system, demonstrating its ability to adapt to and model time-varying instrumental effects by predicting the full three-dimensional Noise Parameter surfaces. By processing the sequence of observations, the model effectively captured the underlying sinusoidal variation in $T_{\text{min}}$ within its predicted surface, whilst simultaneously determining the surfaces for the other stable Noise Parameters and the gain. It is worth noting here that there is some degeneracy between $T_{\text{min}}$ and the other noise parameters, causing a light ripple to appear in the time axis of the other heatmaps. This does not affect the overall calibration result.

Figure~\ref{fig:temporal_heatmap} illustrates the network's performance in tracking the evolving Noise Parameters over the simulated observation period. The heatmaps display the inferred values for each Noise Parameter as a function of time. Notably, the heatmap corresponding to $T_{\text{min}}$ clearly shows the sinusoidal oscillation in the time axis introduced into the simulation.

This result signifies an important extension of radiometer calibration into the temporal domain, potentially for the first time. Our machine learning framework moves beyond static, two-dimensional system characterisation (parameter vs. frequency) to predict three-dimensional Noise Parameter surfaces (parameter vs. frequency and time), learning how the physical properties of the instrument evolve. This capability to model temporal variations is particularly valuable as radio astronomy pushes towards deploying instruments in increasingly remote and challenging environments, both on Earth and in space (e.g., lunar missions like LuSEE-Night~\cite{bale2023lusee} or potential orbital concepts such as CosmoCube~\cite{artuc2024spectrometer}).

\section*{Data availability}
The data that support the findings of this study are available from the corresponding author upon reasonable request.

\section*{Code availability}
The code that supports the findings of this study is available from the corresponding author upon reasonable request.

\section*{Competing interests}
The authors declare no competing interests.

\bibliography{sn-bibliography}

\end{document}